\documentclass{article}
\usepackage[utf8]{inputenc}
\usepackage{graphicx}	
\usepackage{subfigure}
\usepackage[authoryear]{natbib}
\usepackage{stfloats,mathrsfs}
\usepackage{amsmath}

\title{New data for the LBV in NGC\,4736}
\author{
Y. Solovyeva $^1$,\thanks{E-mail:solovyeva@sao.ru}
A. Vinokurov $^1$,
A. Kostenkov $^1$,
A. Sarkisyan$^1$,\\
K. Atapin$^2$,
A. Valeev$^1$
 }
 \date{ 
 $^1$Special Astrophysical Observatory, Nizhnij Arkhyz, 369167, Russia;\\[2ex]%
 $^2$Sternberg Astronomical Institute, Lomonosov Moscow State University, Universitetskij Pr. 13, Moscow 119992, Russia\\[2ex]%
 March 2020}
\providecommand{\keywords}[1]{\textbf{\textit{Key words:}} #1}
\providecommand{\acknowledgements}[1]{\textbf{\textit{Acknowledgements:}} #1}

\begin{document}
\maketitle

\begin{abstract}
We present new spectral and photometric data of confirmed LBV star from the NGC\,4736 galaxy. The star NGC\,4736\_1 ($M_{bol} \approx -11.5^m$) showed noticeable  spectral variability from 2015 to 2018, which was accompanied by a significant change in the brightness. We also have estimated possible initial mass of the object NGC\,4736\_1 as $\sim 130$ M$\odot$.
\end{abstract}
\section{Introduction}
Luminous blue variables (LBVs) are rare type of massive stars with masses $M > 25 M_{\odot}$ \citep{Humphreys2016} at one of their final evolutionary stages. These stars are characterized by a high luminosity of about $\sim10^5 - 10^6 L_{\odot}$ \citep{Humphreys1994} and significant spectral and photometric variability at different time scales.

We present new spectral and photometric data on the LBV star NGC\,4736\_1 \citep{Solovyeva19} (12:50:57.264, +41:07:23.13) from the NGC 4736 galaxy (distance modulus of $m-M = 28.31$, \cite{Tully2013}).

\section{Data}

The first spectra of the NGC\,4736\_1 were obtained with SCORPIO focal reducer \citep{Afanasiev05} on the BTA telescope of the SAO RAS on 2015/01/18 (VPHG1200G grism). NGC\,4736\_1 was also observed on 2018/02/18 (VPHG1200G grism) and 2020/01/18  (VPHG1200B grism). New photometric data were obtained with the BTA (2020) of SAO RAS simultaneously with spectroscopy and Zeiss-1000 (2019/04/09) of SAO RAS and also with 2.5-m telescope of the Caucasian Mountain Observatory of SAI MSU (2020/03/07). We added these data to light curve from previous work \citep{Solovyeva19} (see Fig. \ref{Fig2}, left).

\section{The spectral variability}

The object NGC\,4736\_1 showed significant photometric variability: \linebreak $\Delta V = 1.18^m \pm 0.12^m$ and $\Delta B = 0.90^m \pm 0.12^m$ in V and B bands \citep{Solovyeva19}. The measurements with minimal errors were used for calculations. Based on photometric variability, spectra and luminosity NGC\,4736\_1 was concluded to be an LBV star.

\begin{figure}[!h] 
\centering
\includegraphics[angle=270,scale=0.50]{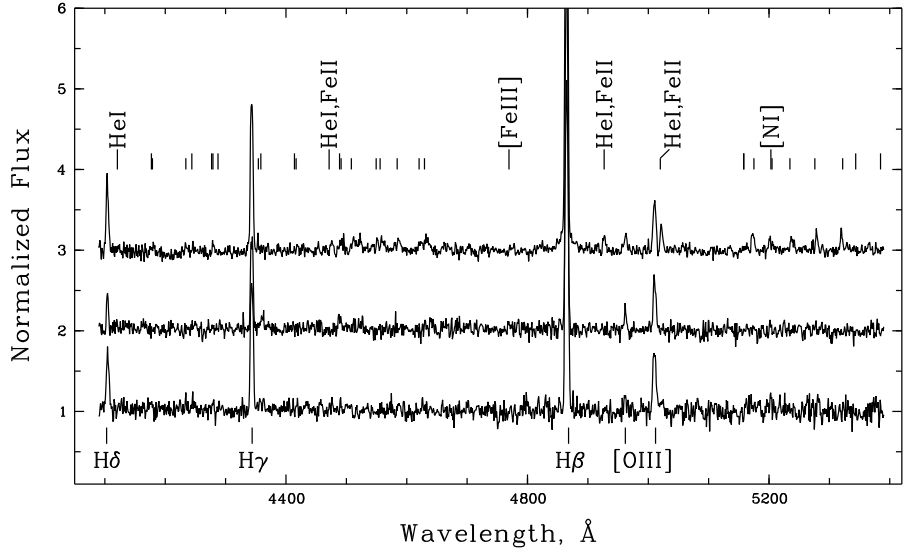} 
\caption{Spectra of NGC\,4736\_1 , obtained on 2015 (top), 2018 (middle) and 2020 (bottom). The main spectral lines are indicated. The unlabelled short and long ticks designate the Fe\,II and [Fe\,II] lines, respectively.} 
\label{Fig1} 
\end{figure}

The spectra of NGC\,4736\_1, obtained in 2015, shows the typical features of well-known LBV stars: broad and bright hydrogen lines, He\,I lines, many iron Fe\,II lines (figure \ref{Fig1}) and a large number of forbidden iron lines [Fe\,II], [Fe\,III]. The spectrum obtained in 2018 has significant changes compared with the spectrum obtained in 2015: the Fe\,II and He\,I lines became indistinguishable, and the broad component of the H$\beta$ line significantly weakened.
The changes in the spectrum are accompanied by a decrease in brightness by $\approx1$ mag in V band from 2015 to 2018. Such behavior means an increase in the photosphere temperature (see Figure 2 in \citealt{Solovyeva19}), which is typical for LBV stars. 

\begin{figure}[!h]
\begin{center} 
\includegraphics[angle=0,scale=0.49]{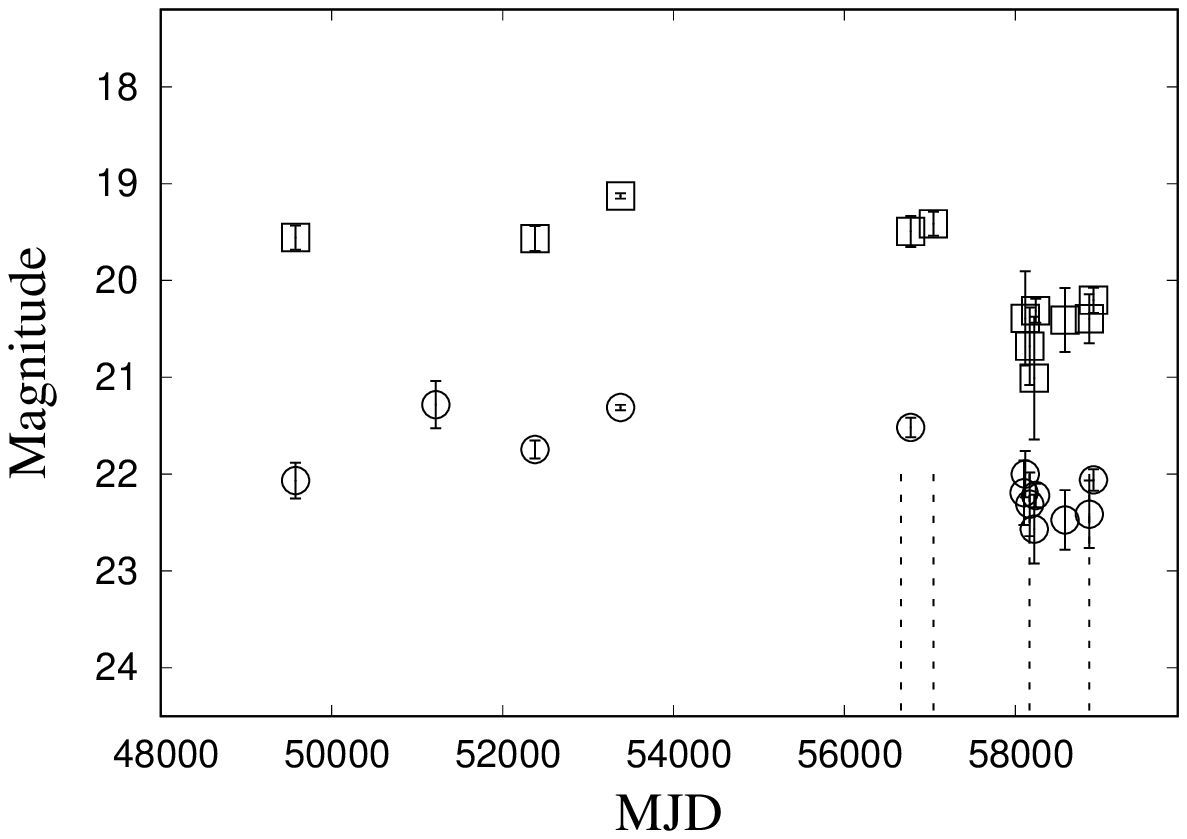} 
\includegraphics[angle=0,scale=0.38]{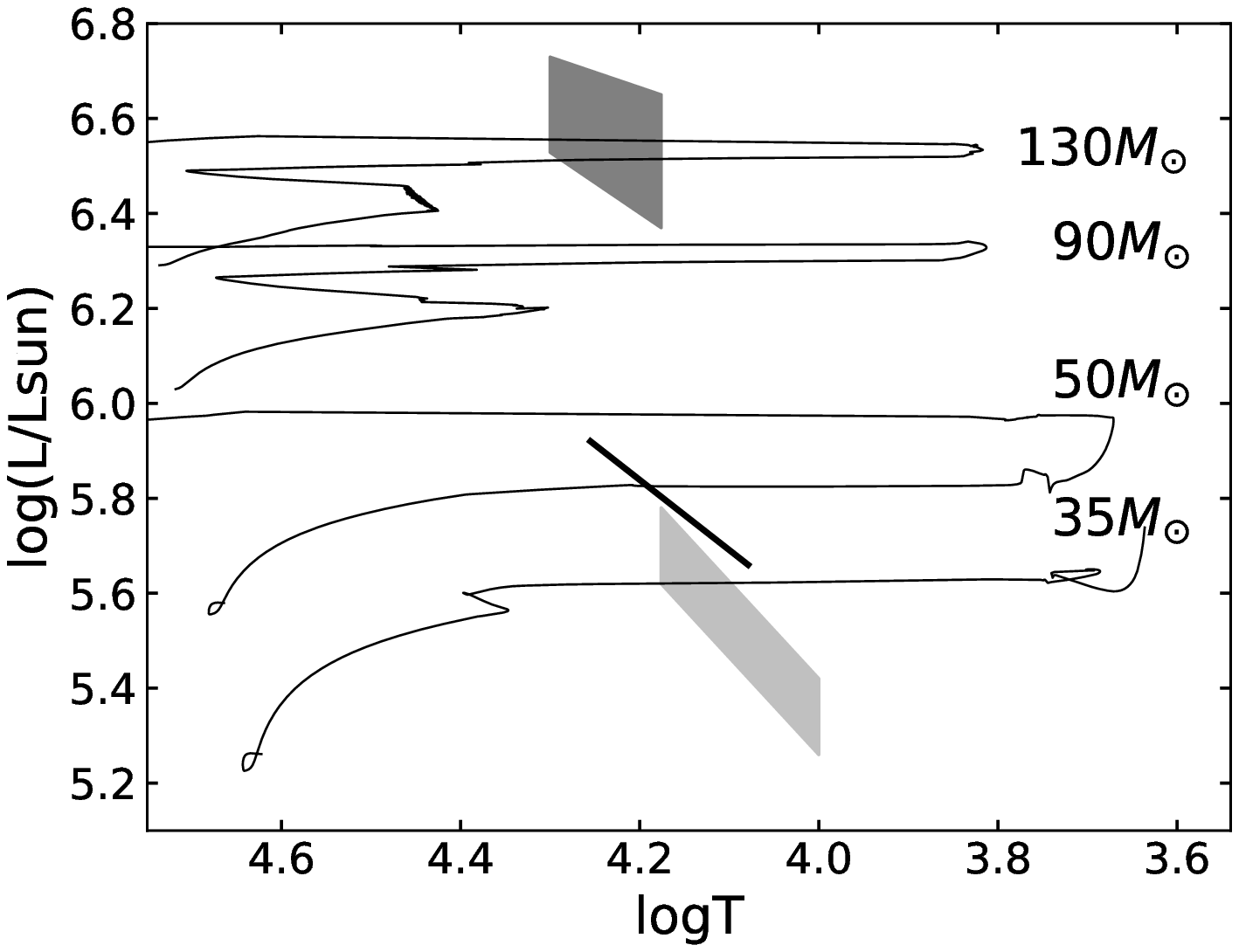} 
\caption{(left panel) Light curve of NGC 4736 1 in the B (circles) and V (squares) bands. The B band is shifted down by two magnitudes. (right panel) Temperature - luminosity diagram for NGC\,4736\_1, NGC\,4736\_2, NGC\,4736\_3 (from top to bottom). Possible ranges of temperature and bolometric luminosity are indicated by gray figures (NGC\,4736\_1, NGC\,4736\_3) and a black solid line (NGC\,4736\_2). Stellar evolutionary tracks from \cite{Tang} for massive stars were used.} 
\label{Fig2} 
\end{center} 
\end{figure}

The spectrum obtained in 2020 does not show a changes compared to the spectrum obtained on 2018. Moreover, the photometric variability was also not detected from 2018 to 2020 (see figure \ref{Fig2}, left panel).

The figure \ref{Fig2} (right panel) shows the position of NGC\,4736\_1 on the temperature - luminosity diagram. The positions of NGC\,4736\_2, NGC\,4736\_3 from \citep{Solovyeva19} are also shown. We used stellar evolutionary tracks from \cite{Tang} with $Z = 0.01$ for NGC\,4736 \citep{Pilyugin}.  Gray areas show the range of possible temperature and luminosity values of objects. The photosphere temperatures and bolometric luminosities estimates, given in \cite{Solovyeva19}, were used.

NGC\,4736\_1 has initial mass of about 130 $M_{\odot}$, while NGC\,4736\_2 and  \linebreak NGC\,4736\_3 may have an initial mass of about 50 and 35 $M_{\odot}$. A mass of 40 $M_{\odot}$ is enough  to pass the LBV stage \citep{Maeder1996}, so the star NGC\,4736\_3 is less luminous LBV candidate.

\acknowledgements{
The reported study was funded by RFBR according to the research project N 19-52-18007.  The authors acknowledge partial support from M.V.Lomonosov Moscow State University Program of Development.
Observations with the SAO RAS telescopes are supported by the Ministry of Science and Higher Education of the Russian Federation (including agreement No05.619.21.0016, project ID RFMEFI61919X0016).}

\end{document}